%% file: gsc.tex
\documentclass[conference,letterpaper]{IEEEtran}

\usepackage[cmex10]{amsmath} 

\usepackage{enumerate}
\usepackage{svg}
\usepackage{tikz}
\usepackage{makecell}
\usepackage[flushleft]{threeparttable}
\usepackage{numprint}
\npthousandsep{,}

\newtheorem{definition}{Definition}
\usepackage[utf8]{inputenc} 
\usepackage[T1]{fontenc}
\usepackage{accents}
\usepackage{url}
\usepackage{ifthen}
\usepackage{cite}

\usepackage{amsfonts}
\usepackage{amssymb}
\newcommand\abs[1]{\left\lvert #1 \right\rvert}
\begin{document}
\include{com}

\title{Generalized Step-Chirp Sequences With Flexible Bandwidth} 


\author{%
  \IEEEauthorblockN{Cheng Du, Yi Jiang}
  \IEEEauthorblockA{Department of Communication Science and Engineering\\
                    Fudan University\\
                    Shanghai, China\\
                    Email: cdu15@fudan.edu.cn, yijiang@fudan.edu.cn}
}


\maketitle


\begin{abstract}
   Sequences with low aperiodic autocorrelation sidelobes have been extensively researched in literatures. With sufficiently low integrated sidelobe level (ISL), their power spectrums are asymptotically flat over the whole frequency domain. However, for the beam sweeping in the massive multi-input multi-output (MIMO) broadcast channels, the flat spectrum should be constrained in a passband with tunable bandwidth to achieve the flexible tradeoffs between the beamforming gain and the beam sweeping time. Motivated by this application, we construct a family of sequences termed the generalized step-chirp (GSC) sequence with a closed-form expression, where some parameters can be tuned to adjust the bandwidth flexibly. In addition to the application in beam sweeping, some GSC sequences are closely connected with Mow's unified construction of sequences with perfect periodic autocorrelations, and may have a coarser phase resolution than the Mow sequence while their ISLs are comparable.
\end{abstract}

\section{Introduction}
Sequences with low aperiodic autocorrelation sidelobes are desirable in communications and radar engineering, e.g., some of the chirp-like sequences developed in \cite{frank1963polyphase, chu1972polyphase, milewski1983periodic, popovic1992generalized, mow1996new}. With a low integrated sidelobe level (ISL), these sequences have quite flat spectrums \cite{schmidt2013problem}, which can be utilized to achieve omnidirectional precoding in broadcast channels \cite{MengXiaGao2018}. 

In 5G NR broadcast channels, the discrete Fourier transform (DFT) codebook is adopted for broadcasting common messages \cite[Section 6.1.6.3]{TS38.802}  in the initial stage of communication. With the energy concentrated in the pointing direction, the maximum beamforming gain can be achieve by the DFT codebook. But for future wireless communication systems with massive number of antennas, the resultant beam would be too narrow, thus requiring many times of beam sweeping to cover the whole angular domain. In contrast, the chirp-like sequence-based omnidirectional beamforming spreads the energy in the whole angular domain, and thus avoids beam sweeping and improves the time efficiency. The omnidirectional beamforming, however, has no beamforming gain, and therefore may have insufficient range coverage for the millimeter-wave or terahertz-wave communication systems where a high beamforming gain is required for compensating the severe path loss.

To circumvent such a dilemma, it is desirable to achieve flexible tradeoffs between the beamforming gain and the beam sweeping time, as pursued by the 3GPP \cite{3gpp.R1.1611929}. From the aspect of spectrum, we aim at designing sequences whose power variation in the passband and power leakage in the stopband should be as small as possible, and the bandwidth of the passband should be flexibly tunable. Besides, their entries should have equal amplitudes for maximizing the energy efficiency of power amplifiers (PAs), and their phase resolutions should be coarse for the implementation using a low-cost phase shifter network (PSN). 

Literatures on this topic include some numerical optimizations \cite{rowe2014spectrally, sergeev2017enhanced, ma2022passive} and some schemes with closed-form solutions \cite{3gpp.R1.1611929, xiao2016hierarchical, xiao2018enhanced, fonteneau2021systematic, du2023hierarchical}. Compared with the numerical optimizations, the schemes with closed-form solutions are easier for hardware implementation, but the bandwidth is less flexible except for the scheme in \cite{fonteneau2021systematic}. The sequence inferred from \cite{fonteneau2021systematic}, referred to as the generalized chirp (GC) sequence in this paper, has flexible bandwidth the same as the numerical counterparts, and its spectrum in the passband is asymptotically flat \cite{fonteneau2021systematic}. Nevertheless, for the GC sequence, the phase resolution of the PSN is too fine to be cost-effective when the number of antennas is large, as shown in our simulations. 

In recent years, polyphase sequences with low correlations and spectrally-null constraints were constructed in \cite{hu2014sequence, liu2018spectrally, tian2020family, ye2022new}, whose $N$-point spectrums (with $N$ being the sequence length) are ideally flat in the passbands and are ideally null in the stopbands. Nevertheless, the $N$-point spectrum is insufficient for beamforming because the user equipments (UEs) are distributed in a continuous angular range, rather than the $N$ discrete directions. Besides, the passbands are interleaved with the stopbands \cite{ye2022new} and the bandwidths are less flexible. Hence, they are still not suitable for beam sweeping.

To achieve flexible tradeoffs between the beamforming gain and the beam sweeping time, in this paper we construct a family of polyphase sequences with flexible bandwidth, termed as the generalized step-chirp (GSC) sequence. The GSC sequence enjoys a coarser phase resolution than the GC sequence. Besides, when the passband stretches over the whole frequency domain, the GSC sequence degenerates into a low-ISL sequence closely connected with the Mow sequence \cite{mow1996new} with perfect periodic autocorrelation, and may require a coarser phase resolution than the Mow sequence. 

Notations: $\lfloor \cdot \rfloor$ stands for taking the floor value. ${\mathbb Z}^+$ represents the set of positive integers, $\Znum_n = \{0, 1, \cdots, n-1\}$. $\omega_N = e^{j\frac{2\pi}{N}}$. $\lVert\cdot\rVert$ is the Frobenius norm. For $x, y, s, t \in \Rnum$, $x\equiv y \mod  s$ stands for that $x-y$ is an integer multiple of $s$; $x = y \mod [s, t)$ means that $x-y$ is an integer multiple of $\abs{s-t}$ and $y \in [s, t)$; $x = y \mod  s$ is equivalent to $x = y \mod [0, s)$.

\section{Preliminaries}
In this section, we review two kinds of passive beamformings for the common message broadcasting: the conventional beam sweeping based on the DFT codebook \cite[Section 6.1.6.3]{TS38.802} in Section \ref{subsec:DFT} and the omnidirectional beamforming based on the chirp-like sequence \cite{mow1996new} in Section \ref{subsec:perfect}.
\subsection{DFT Codebook-based Beam Sweeping} 
\label{subsec:DFT}

Consider a uniform linear array (ULA) of $N$ isotropic antennas with half wavelength spacing. Given a beamforming vector $\abf = [a_0, a_1, \cdots, a_{N-1}]$ with $\abs{a_n} = \frac{1}{\sqrt{N}}, n\in \Znum_N$, the radiated power at azimuth angle $\theta$ and elevation angle $\varphi$ is 
\ben \label{eq:rec}
y(u) = \abs{\sum_{n=0}^{N-1} a_{n} e^{-j\pi nu}}^{2}
\een
where $u = \cos\varphi\cos\theta$. Note that $-1\leq u \leq 1$, hence $y(u)$ is essentially the power spectrum of the sequence $\abf$.

A DFT codeword is $\dbf(u_0) = [d_0, d_1, \cdots, d_{N-1}]$ with $d_n = \frac{1}{\sqrt{N}} e^{j\pi nu_0}, n\in \Znum_N$, where $u_0$ is the beam direction in the $u$-domain. Let $\Delta u \triangleq u- u_0$, then the radiated power is 
\ben \label{eq:dft_pow}
y(u) = \frac{1}{N}\abs{\sum_{n=0}^{N-1}e^{-j\pi n\Delta u}}^2 = \begin{cases}
\abs{\frac{\sin\left(\frac{\pi N}{2}\Delta u\right)}{\sqrt{N}\sin\left(\frac{\pi }{2}\Delta u\right)}}^2, &\Delta u \neq 0\\
N, &\Delta u = 0
\end{cases}.
\een
By \eqref{eq:dft_pow}, the maximum beamforming gain (the ratio of the maximum received power to the average received power) $N$ can be achieved if $u = u_0$, and for a sufficiently large $N$,
\ben \label{eq:half_power_width}
\lim_{N \to \infty} \frac{y\left(u_0\pm\frac{1}{N}\right)}{y(u_0)} = \abs{\lim_{N \to \infty} \frac{1}{N \sin\left(\frac{\pi}{2N}\right)}}^2 = \frac{4}{\pi^2} \approx 0.4,
\een
i.e., $u_0\pm\frac{1}{N}$ is closed to the half-power points of the beam. Hence the DFT codeword $\dbf(u_0)$is designed to cover $[u_0-\frac{1}{N}, u_0+\frac{1}{N}]$. Then a DFT codebook $\{\dbf(u_0)\ \vert \ u_0 \in \mathcal{T}\}$ with $\mathcal{T} = \{\frac{2i+1}{N} -1\vert i\in \Znum_N\}$ is adopted to sweep the beam over the whole space for broadcasting common message, as illustrated by Fig. \ref{fig:dft_vs_mow} (a). The beam sweeping would consume too many time slots if $N$ is large.


\subsection{Chirp-like Sequence-based Omnidirectional Beamforming} \label{subsec:perfect}
In contrast to the beam sweeping that requires many time slots, the omnidirectional beamforming aims at broadcasting messages using only one time slot, which can be achieved by designing a sequence with a flat power spectrum. 

\begin{definition}
    For a length-$N$ complex sequence $\abf$ with $\lVert \abf \rVert = 1$, its aperiodic autocorrelation is defined as
\ben \label{eq:acorr}
R_{a}(\tau) \triangleq \sum_{n=0}^{N-1}a_n \overline{a}_{n-\tau}, \quad 1-N\leq \tau \leq N-1,
\een
where $a_n=0$ if $n < 0$ or $n \geq N$, and the overbar represents the complex conjugation. 
\end{definition}

The power spectrum of $\abf$ is 
\ben
y(u) = \sum_{\tau=1-N}^{N-1} R_{a}(\tau) e^{-j\pi u\tau},
\een
and the variance of the power spectrum is 
\bea \label{eq:continuous}
&\frac{1}{2}\int_{-1}^{1} \left( y(u) - 1\right)^2 \,du\\
=& \frac{1}{2} \sum_{\tau_1 \neq 0}\sum_{\tau_2 \neq 0} R_{a}(\tau_1) R_{a}^*(\tau_2) \int_{-1}^{1}  e^{-j\pi u(\tau_1-\tau_2)}\,du\\
=& \sum_{\tau \neq 0} \abs{R_{a}(\tau)}^2 \triangleq ISL_{a}
\eea
where $ISL_{a}$ is the integrated sidelobe level (ISL) of $R_{a}(\tau)$. Hence for omnidirectional beamforming, the sequence 's ISL should be small, e.g., some of the chirp-like sequences \cite{frank1963polyphase, chu1972polyphase, milewski1983periodic, popovic1992generalized, mow1996new}. As a unified construction of sequences with perfect periodic autocorrelation, the Mow sequence family \cite{mow1996new} where some sequences have low ISL, is given below for ease of reference.


\begin{definition} \label{def:mow}\cite{mow1996new}
    The Mow sequence is a kind of sequences of length $N = sm^2$ with $s$ being the square-free part of $N$, whose entries are $\frac{1}{\sqrt{N}}\omega_N ^{m\xi_n}, n \in \Znum_N$  with
\ben \label{eq:mow}
\xi_{km+l} = mc(s)\alpha(l)k^2 + \beta(l)k + f_l(0),\ l\in \Znum_m, k \in \Znum_{sm}
\een
where 
\ben
c(s) = \begin{cases}
    \frac{1}{2}, & {\rm for}\ s\ {\rm even}\\
    1, & {\rm for}\ s\ {\rm odd}
\end{cases}\ ,
\een
$\alpha(l) \in \{1, 2, \cdots, s-1\}$ is any function with $\gcd(\alpha(l), s)=1, \forall l \in \Znum_m$, and  $\beta(l) \in \Znum_{sm}$ is any function such that $\beta(l)\ \text{mod}\ m$ is a permutation of $\Znum_m$, and $f_l(0), \forall l \in \Znum_m$ are any rational numbers. 
\end{definition}

\begin{figure}[ht!]
\centering
\includegraphics[width=.48\textwidth]{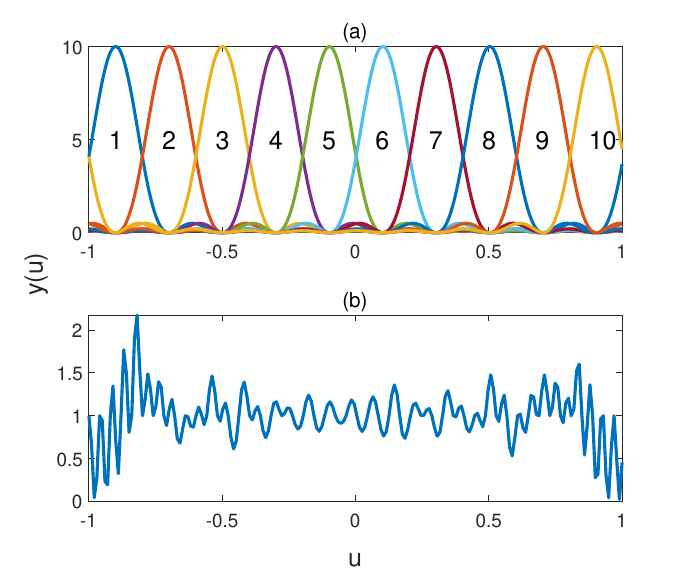}
\caption{The power spectrums of two kinds of beamformings. (a): the DFT codebook-based beam sweeping; (b) the chirp-like sequence-based omnidirectional beamforming.}
\label{fig:dft_vs_mow}
\end{figure}

The beam sweeping based on the DFT codebook and the omnidirectional beamforming based on the Mow sequence are compared in Fig. \ref{fig:dft_vs_mow}. For the DFT codebook, $N=10$; for the Mow sequence, $N=50$, $s=2$, $m=5$, $c(s) = \frac{1}{2}$, $\alpha(l)=1$, $\beta(l) = l-25$, $f_l(0) = -9.5l$. The DFT codebook in Fig. \ref{fig:dft_vs_mow} (a) achieves the maximum beamforming gain but requires $10$ times of beam sweeping, while the Mow sequence in Fig. \ref{fig:dft_vs_mow} (b) can broadcast messages in one time slot but has no beamforming gain.

\section{Generalized Step-chirp Sequence} \label{sec:step_chirp}
To achieve flexible tradeoffs between the beamforming gain and the beam sweeping time, in Section \ref{subsec:gsc}, we construct a family of polyphase sequences with tunable bandwidth, termed as the generalized step-chirp (GSC) sequence; in Section \ref{subsec:relation}, we discuss the relationships between the GSC sequence, the DFT codebook, the generalized chirp (GC) sequence inferred from \cite{fonteneau2021systematic} and the Mow sequence \cite{mow1996new}.

\subsection{Construction of Generalized Step-chirp Sequence} \label{subsec:gsc}
\begin{figure}[ht!]
\centering
\includegraphics[width=.48\textwidth]{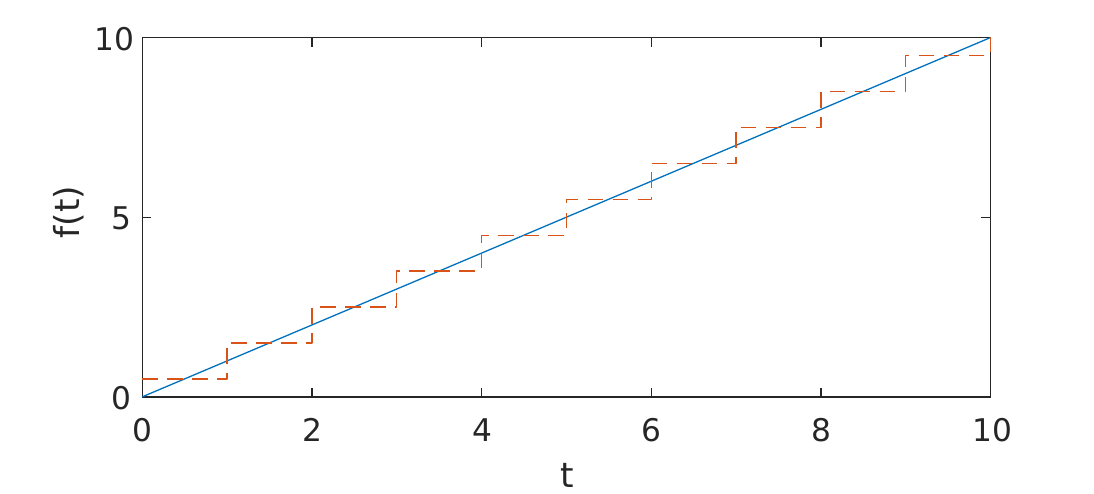}
\caption{Step approximation of LFM, $a=1, b=\frac{1}{2}$, and $T=10$.}
\label{fig:step_fre}
\end{figure}

Consider a step-chirp signal as follows:
\ben \label{eq:step_chirp}
c(t) = e^{j2\pi \int_{0}^{t}f(\tau) d\tau}, \quad 0\leq t \leq T,
 \een
where $f(t)$ is a step approximation of linear frequency modulation (LFM):
\ben
f(t) = a(\lfloor t \rfloor +b), \ 0\leq t\leq T
\een
for $\forall a > 0, \forall b\in \Rnum$. An LFM $f^{\prime}(t) = t$ and its step approximation $f(t)$ with $a=1, b=\frac{1}{2}$, and $T=10$ are illustrated by Fig. \ref{fig:step_fre}. The bandwidth of the step-chirp signal is $aT$ approximately. Besides, we require the Nyquist sampling number $aT^2\geq 1$. 

Now sample $c(t)$ in (\ref{eq:step_chirp}) at rate $m \triangleq aT/\gamma$ with $0<\gamma \leq 1$, where $m$ is assumed to be an integer via setting $a$ properly. We then obtain $N$ samples 
\begin{equation} \label{eq.ctnm}
    c\left(t \right)|_{t=\frac{n}{m}} = e^{j\phi_n},\quad n \in \Znum_N
\end{equation}
with 
\ben  \label{eqNmT}
 N = mT = aT^{2}/\gamma = m^2\gamma /a.
\een 
Because $aT^2\geq 1$, we have $\gamma = \frac{aT^2}{N} \geq \frac{1}{N}$.

Factor $n\in \Znum_N$ into
\ben
n = km+l, \ k\triangleq\left\lfloor n/m \right\rfloor, \ l \in \Znum_m.
\een
Direct calculations show that
\ben 
\phi_n = 2\pi \frac{ak\left(k-1+2b\right)}{2}+2\pi \frac{a\left(k+b\right)l}{m}. 
\label{eq:step_chirp_sample2}
%
\een
Note from \eqref{eqNmT} that $a=\frac{m^2\gamma}{N}$; thus, 
\ben \label{eq:step_chirp_sample}
 \phi_n = \frac{2\pi}{N} m\gamma\left( \frac{k(k-1)}{2}m + kl + bn\right).
\een


Besides, the Fourier transform of $c(t)$ can be derived to be a weighted summation of $T$ sinc functions:
\ben
C(f) = \sum_{i=0}^{T-1} e^{j\pi [ai^2 + 2(ab-f)i+ab-f]}{\rm sinc} [f-a(i+b)]. 
\een
At $f_0 \triangleq a(b-\frac{1}{2})$, the value of the left-most sinc function ($i=0$) is $\text{sinc}(-\frac{a}{2})$; at $f_1 \triangleq a(b-\frac{1}{2}+T)$, the value of the right-most sinc function ($i=T-1$) is $\text{sinc}(\frac{a}{2})$. Hence the interval $(-\infty, f_0) \cup (f_1, +\infty)$ can be regarded as the stopband since most of the sinc functions have attenuated to a low level. Because the bandwidth of the step-chirp signal is $aT$ approximately, the interval $[f_0, f_1]$ can be regarded as the passband of $C(f)$. Note that the analog bandwidth $aT$ is scaled to be the digital bandwidth $2\pi\gamma$ by over-sampling, hence the passband of the sample sequence is $\left[\omega_0, \omega_0 +2\pi\gamma\right]$ where 
\ben
\omega_0 = \frac{2\pi\gamma}{aT}f_0  = \frac{2\pi}{N}m\gamma\left(b-\frac{1}{2}\right).
\een


The above arguments established the following theorem.
\begin{theorem} \label{thm:gsc}
The GSC sequence is a family of polyphase sequences with entries $\frac{1}{\sqrt{N}}\omega_N^{m\zeta_n}, n \in \Znum_N$, where
\bea \label{eq:gsc}
\zeta_{n} =& \gamma\left( \frac{k(k-1)}{2}m + kl + bn\right),\\
& k= \left\lfloor n/m \right\rfloor,\ l = n-km,
\eea
with parameter set $\{N, \gamma, m, b\ \vert\  N \in \Znum^+, m|N, \frac{1}{N}\leq\gamma\leq1, b\in \Rnum\}$.
The passband of the power spectrum of the GSC sequence is  
\ben
   \left[\omega_0, \omega_0 +2\pi\gamma\right]
\een
where $\omega_0 = \frac{2\pi}{N}m\gamma\left(b-\frac{1}{2}\right)$. For beam sweeping, the beam is pointed at $u_0$ to cover $[u_0-\gamma, u_0 + \gamma)$, where 
\ben \label{eq:ug}
u_0 = \frac{2}{N}m\gamma\left(b-\frac{1}{2}\right) + \gamma \ \text{mod}\  [-1, 1).
\een
\end{theorem}

The bandwidth of the GSC sequence can be flexibly adjusted by tuning the parameter $\gamma$, thus achieving the flexible tradeoffs between the beamforming gain and the beam sweeping time, as shown in Simulations.

\subsection{Relationships Between the GSC sequence and Other Sequences} \label{subsec:relation}
\begin{figure}[ht!]
\centering
\includegraphics[width=0.4\textwidth]{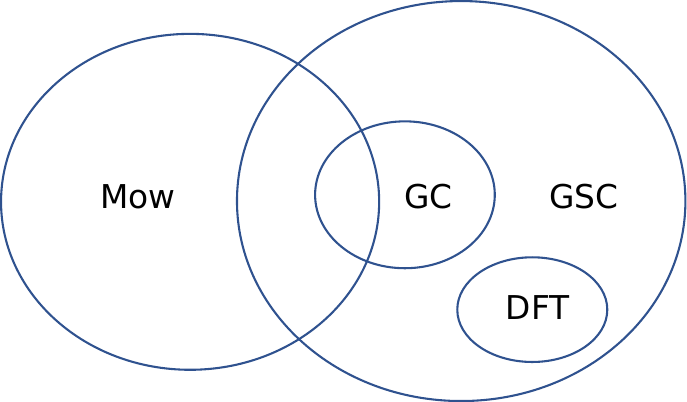}
\caption{The relationships between the Mow sequence, the GC sequence, the DFT codebook and the GSC sequence.
\label{fig:family}}
\end{figure}

The relationships between the GSC sequence, the DFT codebook, the GC sequence and the Mow sequence family are illustrated by Fig. \ref{fig:family}, as explained below.

\subsubsection{GSC Sequence and DFT Codebook}
    In Theorem \ref{thm:gsc}, let $m=N$ and $\gamma = \frac{1}{N}$. Then we have $k=0$ and $n=l$ for $\forall \ n\in \Znum_N$. This degenerated GSC sequence has entries
    \ben
    \frac{1}{\sqrt{N}}e^{j\frac{2\pi}{N}m \zeta_n} = \frac{1}{\sqrt{N}}e^{j\frac{2\pi}{N}bn}, \ n\in \Znum_N,
    \een
    and the passband is 
    \ben
    \left[ \frac{2\pi}{N}\left(b-\frac{1}{2}\right),\ \frac{2\pi}{N}\left(b+\frac{1}{2}\right)\right].
    \een
    According to Section \ref{subsec:DFT}, this is exactly a DFT codeword pointing at $u_0 = \frac{2b}{N}\ \text{mod}\ [-1, 1)$. Hence the GSC sequence encompasses the DFT codebook and thus may be backward-compatible with the current industrial standard.

\subsubsection{GSC Sequence and GC Sequence}
    When $m=1$, we have from \eqref{eq:gsc} that $\frac{1}{\sqrt{N}}\omega_N^{m\zeta_n} = \frac{1}{\sqrt{N}}\omega_N^{\gamma \frac{n(n+2b-1)}{2}}$, which is the GC sequence inferred from over-sampling a chirp signal \cite{fonteneau2021systematic}. Therefore, the GSC sequence is also a generalization of the GC sequence. The phase resolution of a sequence with phases in $\{\frac{2\pi p}{P}\vert p\in \Znum_P\}$ is $\frac{2\pi}{P}$. Note that the parameter $m$ can be tuned for coarser phase resolution, e.g., suppose $b\in \Znum$ and $\gamma$ is a rational number of form $\frac{p}{q}$ with $p, q$ coprime, then the phase resolutions are
    \ben
    R_{gsc} = \frac{2\pi}{Nq / \text{gcd}(Nq, mp)}, \ 
    R_{gc} = \frac{2\pi}{Nq / \text{gcd}(Nq, p)},
    \een 
    from which we have $R_{gc} \leq R_{gsc} \leq m R_{gc}$, e.g., if $p=1$, then $R_{gsc} = mR_{gc}$.

\subsubsection{GSC Sequence and Mow Sequence}
Set $\gamma = 1$ (i.e., the Nyquist sampling), and we obtain another kind of degenerated GSC sequence with entries $\frac{1}{\sqrt{N}}\omega_N^{m\zeta_n}, n \in \Znum_N$, where
\bea \label{eq:gsc_subset}
\zeta_{n} =&\ \frac{k(k-1)}{2}m + kl + bn,\\
& k= \left\lfloor n/m \right\rfloor,\ l = n-km,
\eea
with parameter set $\{N, m, b\, \vert\, N\in \Znum^+, m|N, b\in \Rnum\}$,
which is related to the Mow sequence as shown below. 
\begin{Proposition} \label{prop:equiv}
With the following two constraints on the parameter $m$ and $b$ in \eqref{eq:gsc_subset}, respectively, the degenerated GSC sequence in \eqref{eq:gsc_subset} is a special case of the Mow sequence in \eqref{eq:mow}:
\begin{enumerate}
    \item $m$ is the square part of $N$, i.e., $N=sm^2$.
    \item $2b$ is odd if $s$ is even or $b$ is an integer if $s$ is odd.
\end{enumerate}
\end{Proposition}

\begin{IEEEproof}
First note that with the first constraint, the sequence length in \eqref{eq:gsc_subset} is $s m^2$, which is the same as the Mow sequence.

Second, if $s$ is even and $2b$ is odd, then \eqref{eq:gsc_subset} is a special case of \eqref{eq:mow} with $c(s) = \frac{1}{2}$, $\alpha(l)=1$, $\beta(l) = \frac{2b-1}{2}m+l$, $f_l(0) = bl$ and $\frac{2b-1}{2}$ is an integer such that $\beta(l) \equiv l\ \text{mod}\ m$ is a permutation of $\Znum_m$.

If $s$ is odd and $b$ is an integer, then denote $s = 2d-1$ for some $d \in {\mathbb Z}^+$. Since $k(k+2b-1)$ is an even number, it holds that 
\ben \label{eq:even_mod}
\begin{aligned}
d k(k+2b-1) &= \frac{k(k+2b-1)}{2} (s+1) \\
&\equiv \frac{k(k+2b-1)}{2}  \ \text{mod}\ s.
\end{aligned}
\een
Rewrite \eqref{eq:gsc_subset} as
\ben \label{eq:step_chirp_sample4}
\zeta_{km+l} =  \frac{k(k+2b-1)}{2}m + kl + bl.
\een
It follows from \eqref{eq:step_chirp_sample4} and \eqref{eq:even_mod} that
\ben
\zeta_{km+l} \equiv d k(k+2b-1)m + kl +bl\ \text{mod}\ sm,
\een
which is a special case of \eqref{eq:mow} with $c(s)=1$, $\alpha(l) = d$ [one may verify that $\gcd(d, 2d-1) = 1$], $\beta(l) = (2b-1)d m + l$, $f_l(0) = bl$.
\end{IEEEproof} 

Indeed, one may relax the constraints in Proposition \ref{prop:equiv} to improve the phase resolution of the degenerated GSC sequence in \eqref{eq:gsc_subset}. The phase resolution of the Mow sequence in \eqref{eq:mow} with $f_l(0)$ being an integer is 
\ben
R_{mow} = \begin{cases}
    \frac{\pi}{N/m}, & \ s\ {\rm is}\ {\rm even}, m\ {\rm is}\ {\rm odd}\\
    \frac{2\pi}{N/m}, &\ {\rm otherwise}
\end{cases},
\een
and the phase resolution of the GSC sequence with $\gamma = 1, b \in \Znum^{+}$ is $R_{gsc} = \frac{2\pi}{N/m}$. If $m$ is larger than the square part of $N$, then the phase resolution of the GSC sequence would be coarser than the Mow sequence as shown in Simulations.

\section{Simulations}
This section presents simulation examples to verify the capability of the GSC sequence in making flexible tradeoffs between the beamforming gain and the beam sweeping time, and its advantages over the GC sequence and the Mow sequence in terms of the phase resolution and the spectrum.

\subsection{Tradeoffs Between the Beamforming Gain and the Beam Sweeping Time}
\begin{figure}[ht!]
\centering
\includegraphics[width=0.48\textwidth]{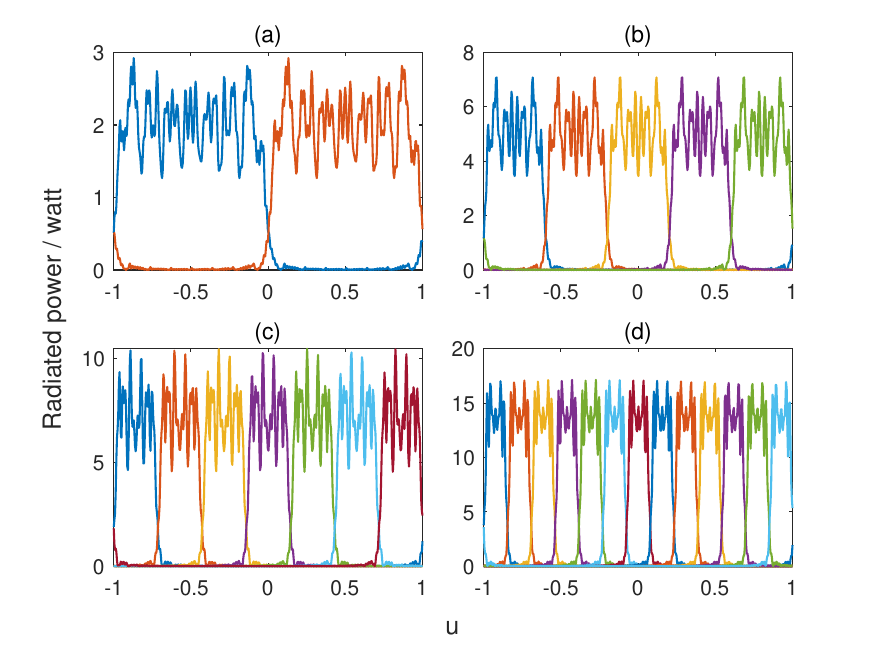}
\caption{Flexible tradeoffs between the beamforming gain and the beam sweeping time. (a): $\gamma = \frac{1}{2}$; (b): $\gamma = \frac{1}{5}$; (c): $\gamma = \frac{1}{7}$; (d): $\gamma = \frac{1}{13}$.}
\label{fig:tradeoff}
\end{figure}

To show the flexibility of the GSC sequence for beam sweeping, we simulate and show in Fig. \ref{fig:tradeoff} the beampatterns of the GSC sequences of length $N=120$ with $(\gamma, m)\in\{(\frac{1}{2}, 15), (\frac{1}{5}, 24), (\frac{1}{7}, 30), (\frac{1}{13}, 40)\}$. The parameter $b$ is chosen so that the beam direction $u_0$ in \eqref{eq:ug} runs through $\{(2i-1) \gamma -1 \vert i=1, 2, \cdots, \frac{1}{\gamma}\}$ for the contiguous coverage of $[-1, 1)$. Fig. \ref{fig:tradeoff} (a) illustrates $2$ times of beam sweeping with 2x beamforming gain while Fig. \ref{fig:tradeoff} (d) represents $13$ times of beam sweeping with 13x beamforming gain. In summary, by adjusting $\gamma$ and $b$ to control the bandwidth and the beam direction, flexible tradeoffs between the beamforming gain and the beam sweeping time can be achieved for efficient beam sweeping. We want to emphasize that the y-axis is in the linear scale. Thus, the power fluctuation in the passband is less than $3$dB.

\subsection{Phase Resolution and Spectrum}
For a GSC sequence $\gbf = [g_0, g_1, \cdots, g_{N-1}]$, the normalized root mean square error (NRMSE) of passband is defined as 
\ben
\sqrt{\frac{1}{\abs{{\cal I}_p}}\sum_{i \in {\mathcal I }_p}\left(\gamma\abs{\sum_{n=0}^{N-1}g_n e^{-j\frac{2\pi}{N^{\prime}}in}}^2 - 1\right)^2}
\een
where $N^{\prime}$ is the DFT length and ${\mathcal I }_p \subset \Znum_{N^{\prime}}$ is the set of passband indices. Here we set $N^{\prime} = 4N$. And the stopband leakage ratio is defined as 
\ben
\frac{1}{N^{\prime}}\sum_{i \in {\mathcal I }_s}\abs{\sum_{n=0}^{N-1}g_n e^{-j\frac{2\pi}{N^{\prime}}in}}^2
\een
where ${\mathcal I }_s = \Znum_{N^{\prime}} \setminus {\mathcal I }_p$ is the set of the stopband indices. 

Compared with the GC sequence and the Mow sequence, the GSC sequence with a proper parameter $m$ may have a coarser phase resolution and a comparable spectrum or even flatter.

\subsubsection{GSC Sequence versus GC sequence}
\begin{figure}[ht!]
\centering
\includegraphics[width=0.48\textwidth]{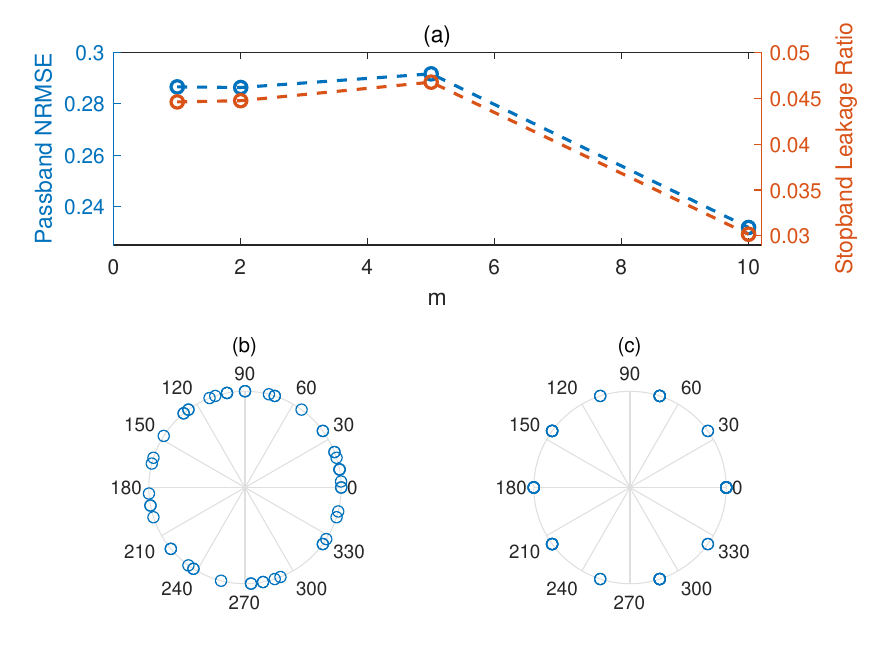}
\caption{The improvement of phase resolution and spectrum of the GSC sequence against the GC sequence. (a): the passband NRMSE and the stopband leakage ratio of a length-$50$ GSC sequence for different $m$; (b): the phases of the GC sequence corresponding to $m=1$ in (a); (c): the phases of the GSC sequence corresponding to $m=10$ in (a).}
\label{fig:phase_improve}
\end{figure}

Fig. \ref{fig:phase_improve} shows the impact of the parameter $m$ on the spectrum and the phase resolution of a GSC sequence, where $N=50$, $\gamma = \frac{1}{2}$, $b=1$. Compared with the GC sequence, i.e., $m=1$, the GSC sequence with $m=10$ has smaller passband NRMSE and stopband leakage ratio as shown in Fig. \ref{fig:phase_improve} (a), and the phase resolution of the proposed GSC sequence is $10$ times coarser as shown in Fig. \ref{fig:phase_improve} (b) and Fig. \ref{fig:phase_improve} (c).

\subsubsection{GSC sequence versus Mow sequence}
\begin{figure}[ht!]
\centering
\includegraphics[width=0.45\textwidth]{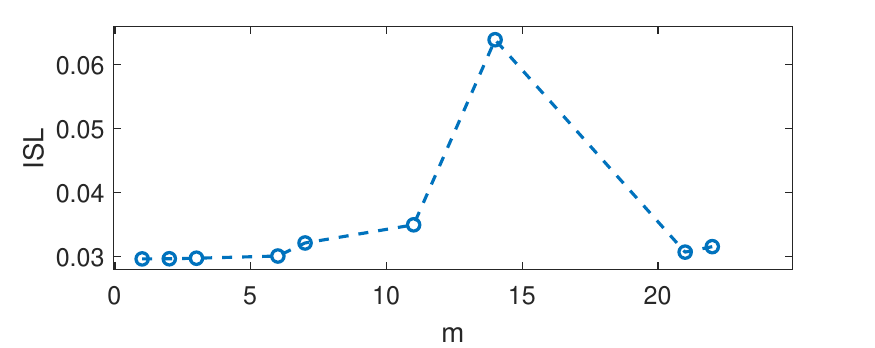}
\caption{The ISL of a length-$462$ GSC sequence for different $m$.}
\label{fig:gsc_mow}
\end{figure}

Fig. \ref{fig:gsc_mow} shows the ISL of a GSC sequence of length $N=462$ for different parameters $m$, with $\gamma = 1$ and $b=\frac{1}{2}$. Note that the square part of $N=462$ is $m=1$, thus the point with $m=1$ in Fig. \ref{fig:gsc_mow} corresponds to a Mow sequence, which can be verified by simulation to have exactly the minimum ISL among all the $55440$ Mow sequences of length $462$ with a phase resolution $\frac{\pi}{462}$ \cite[Theorem 5]{mow1996new}. Remarkably, the ISL for $m=1$ is $0.0297$ and the ISL for $m=21$ is $0.0307$, which means a reduction of phase resolution by a factor of $21$ but with a negligible increase of ISL, i.e., a comparably flat spectrum. 

\section{Conclusions} \label{SEC:con}
In this paper, we construct the generalized step-chirp (GSC) sequence, which can achieve flexible tradeoffs between the beamforming gain and the beam sweeping time for the common message broadcasting in massive MIMO systems. The GSC sequence has a coarser phase resolution than the generalized chirp (GC) sequence, which facilitates its implementation with a low-cost phase shifter network (PSN). Besides, the GSC sequence may have coarser phase resolution than the Mow sequence with a negligible increase of the integrated sidelobe level (ISL).






\bibliographystyle{IEEEtran}
\bibliography{gsc}









\end{document}

%% file: com.tex
\newcommand{\fs}{\hspace{0.07in}}
\newcommand{\bs}{\hspace{-0.1in}}
\newcommand{\re}{{\rm Re} \, }
\newcommand{\e}{{\rm E} \, }
\newcommand{\p}{{\rm P} \, }
\newcommand{\cn}{{\cal CN} \, }
\newcommand{\n}{{\cal N} \, }
\newcommand{\ba}{\begin{array}}
\newcommand{\ea}{\end{array}}
\newcommand{\be}{\begin{displaymath}}
\newcommand{\ee}{\end{displaymath}}
\newcommand{\ben}{\begin{equation}}  
\newcommand{\een}{\end{equation}}
\newcommand{\bea}{\begin{equation}\begin{aligned}}
\newcommand{\eea}{\end{aligned}\end{equation}}      
\newcommand{\bena}{\begin{eqnarray}}
\newcommand{\eena}{\end{eqnarray}}
\newcommand{\beqa}{\begin{eqnarray*}}
\newcommand{\enqa}{\end{eqnarray*}}
\newcommand{\f}{\frac}
\newcommand{\bc}{\begin{center}}
\newcommand{\ec}{\end{center}}
\newcommand{\bi}{\begin{itemize}}
\newcommand{\ei}{\end{itemize}}
\newcommand{\benu}{\begin{enumerate}}
\newcommand{\eenu}{\end{enumerate}}
\newcommand{\bdes}{\begin{description}}
\newcommand{\edes}{\end{description}}
\newcommand{\bt}{\begin{tabular}}
\newcommand{\et}{\end{tabular}}
\newcommand{\vs}{\vspace}
\newcommand{\hs}{\hspace}
\newcommand{\sort}{\rm sort \,}

\newcommand \thetabf{{\mbox{\boldmath$\theta$\unboldmath}}}
\newcommand{\Phibf}{\mbox{${\bf \Phi}$}}
\newcommand{\Psibf}{\mbox{${\bf \Psi}$}}
\newcommand \alphabf{\mbox{\boldmath$\alpha$\unboldmath}}
\newcommand \betabf{\mbox{\boldmath$\beta$\unboldmath}}
\newcommand \gammabf{\mbox{\boldmath$\gamma$\unboldmath}}
\newcommand \deltabf{\mbox{\boldmath$\delta$\unboldmath}}
\newcommand \epsilonbf{\mbox{\boldmath$\epsilon$\unboldmath}}
\newcommand \zetabf{\mbox{\boldmath$\zeta$\unboldmath}}
\newcommand \etabf{\mbox{\boldmath$\eta$\unboldmath}}
\newcommand \iotabf{\mbox{\boldmath$\iota$\unboldmath}}
\newcommand \kappabf{\mbox{\boldmath$\kappa$\unboldmath}}
\newcommand \lambdabf{\mbox{\boldmath$\lambda$\unboldmath}}
\newcommand \mubf{\mbox{\boldmath$\mu$\unboldmath}}
\newcommand \nubf{\mbox{\boldmath$\nu$\unboldmath}}
\newcommand \xibf{\mbox{\boldmath$\xi$\unboldmath}}
\newcommand \pibf{\mbox{\boldmath$\pi$\unboldmath}}
\newcommand \rhobf{\mbox{\boldmath$\rho$\unboldmath}}
\newcommand \sigmabf{\mbox{\boldmath$\sigma$\unboldmath}}
\newcommand \taubf{\mbox{\boldmath$\tau$\unboldmath}}
\newcommand \upsilonbf{\mbox{\boldmath$\upsilon$\unboldmath}}
\newcommand \phibf{\mbox{\boldmath$\phi$\unboldmath}}
\newcommand \varphibf{\mbox{\boldmath$\varphi$\unboldmath}}
\newcommand \chibf{\mbox{\boldmath$\chi$\unboldmath}}
\newcommand \psibf{\mbox{\boldmath$\psi$\unboldmath}}
\newcommand \omegabf{\mbox{\boldmath$\omega$\unboldmath}}
\newcommand \Sigmabf{\hbox{$\bf \Sigma$}}
\newcommand \Upsilonbf{\hbox{$\bf \Upsilon$}}
\newcommand \Omegabf{\hbox{$\bf \Omega$}}
\newcommand \Deltabf{\hbox{$\bf \Delta$}}
\newcommand \Gammabf{\hbox{$\bf \Gamma$}}
\newcommand \Thetabf{\hbox{$\bf \Theta$}}
\newcommand \Lambdabf{\hbox{$\bf \Lambda$}}
\newcommand \Xibf{\hbox{\bf$\Xi$}}
\newcommand \Pibf{\hbox{\bf$\Pi$}}
\newcommand \abf{{\bf a}}
\newcommand \bbf{{\bf b}}
\newcommand \cbf{{\bf c}}
\newcommand \dbf{{\bf d}}
\newcommand \ebf{{\bf e}}
\newcommand \fbf{{\bf f}}
\newcommand \gbf{{\bf g}}
\newcommand \hbf{{\bf h}}
\newcommand \ibf{{\bf i}}
\newcommand \jbf{{\bf j}}
\newcommand \kbf{{\bf k}}
\newcommand \lbf{{\bf l}}
\newcommand \mbf{{\bf m}}
\newcommand \nbf{{\bf n}}
\newcommand \obf{{\bf o}}
\newcommand \pbf{{\bf p}}
\newcommand \qbf{{\bf q}}
\newcommand \rbf{{\bf r}}
\newcommand \sbf{{\bf s}}
\newcommand \tbf{{\bf t}}
\newcommand \ubf{{\bf u}}
\newcommand \vbf{{\bf v}}
\newcommand \wbf{{\bf w}}
\newcommand \xbf{{\bf x}}
\newcommand \ybf{{\bf y}}
\newcommand \zbf{{\bf z}}
\newcommand \rbfa{{\bf r}}
\newcommand \xbfa{{\bf x}}
\newcommand \ybfa{{\bf y}}
\newcommand \Abf{{\bf A}}
\newcommand \Bbf{{\bf B}}
\newcommand \Cbf{{\bf C}}
\newcommand \Dbf{{\bf D}}
\newcommand \Ebf{{\bf E}}
\newcommand \Fbf{{\bf F}}
\newcommand \Gbf{{\bf G}}
\newcommand \Hbf{{\bf H}}
\newcommand \Ibf{{\bf I}}
\newcommand \Jbf{{\bf J}}
\newcommand \Kbf{{\bf K}}
\newcommand \Lbf{{\bf L}}
\newcommand \Mbf{{\bf M}}
\newcommand \Nbf{{\bf N}}
\newcommand \Obf{{\bf O}}
\newcommand \Pbf{{\bf P}}
\newcommand \Qbf{{\bf Q}}
\newcommand \Rbf{{\bf R}}
\newcommand \Sbf{{\bf S}}
\newcommand \Tbf{{\bf T}}
\newcommand \Ubf{{\bf U}}
\newcommand \Vbf{{\bf V}}
\newcommand \Wbf{{\bf W}}
\newcommand \Xbf{{\bf X}}
\newcommand \Ybf{{\bf Y}}
\newcommand \Zbf{{\bf Z}}
\newcommand \Omegabbf{{\bf \Omega}}
\newcommand \Rssbf{{\bf R_{ss}}}
\newcommand \Ryybf{{\bf R_{yy}}}
\newcommand \Cset{{\cal C}}
\newcommand \Rset{{\cal R}}
\newcommand \Zset{{\cal Z}}
\newcommand{\otheta}{\stackrel{\circ}{\theta}}
\newcommand{\defeq}{\stackrel{\bigtriangleup}{=}}
\newcommand{\oabf}{{\bf \breve{a}}}
\newcommand{\odbf}{{\bf \breve{d}}}
\newcommand{\oDbf}{{\bf \breve{D}}}
\newcommand{\oAbf}{{\bf \breve{A}}}
\renewcommand \vec{{\mbox{vec}}}
\newcommand{\Acalbf}{\bf {\cal A}}
\newcommand{\calZbf}{\mbox{\boldmath $\cal Z$}}
\newcommand{\feop}{\hfill \rule{2mm}{2mm} \\}
\newtheorem{theorem}{Theorem}

\newcommand{\Rnum}{{\mathbb R}}
\newcommand{\Cnum}{{\mathbb C}}
\newcommand{\Znum}{{\mathbb Z}}
\newcommand{\Enum}{{\mathbb E}}
\newcommand{\Mnum}{{\mathbb M}}
\newcommand{\Nnum}{{\mathbb N}}
\newcommand{\Inum}{{\mathbb I}}

\newcommand{\Acal}{{\cal A}}
\newcommand{\Bcal}{{\cal B}}
\newcommand{\Ccal}{{\cal C}}
\newcommand{\Dcal}{{\cal D}}
\newcommand{\Ecal}{{\cal E}}
\newcommand{\Fcal}{{\cal F}}
\newcommand{\Gcal}{{\cal G}}
\newcommand{\Hcal}{{\cal H}}
\newcommand{\Ical}{{\cal I}}
\newcommand{\Ocal}{{\cal O}}
\newcommand{\Rcal}{{\cal R}}
\newcommand{\Zcal}{{\cal Z}}
\newcommand{\Xcal}{{\cal X}}
\newcommand{\zzbf}{{\bf 0}}
\newcommand{\zebf}{{\bf 0}}

\newcommand{\eop}{\hfill $\Box$}

\newcommand{\gss}{\mathop{}\limits}
\newcommand{\gs}{\mathop{\gss_<^>}\limits}

\newcommand{\circlambda}{\mbox{$\Lambda$
             \kern-.85em\raise1.5ex
             \hbox{$\scriptstyle{\circ}$}}\,}

\newcommand{\tr}{\mathop{\rm tr}}
\newcommand{\var}{\mathop{\rm var}}
\newcommand{\cov}{\mathop{\rm cov}}
\newcommand{\diag}{\mathop{\rm diag}}
\def\rank{\mathop{\rm rank}\nolimits}
\newcommand{\ra}{\rightarrow}
\newcommand{\ul}{\underline}
\def\Pr{\mathop{\rm Pr}}
\def\Re{\mathop{\rm Re}}
\def\Im{\mathop{\rm Im}}

\def\submbox#1{_{\mbox{\footnotesize #1}}}
\def\supmbox#1{^{\mbox{\footnotesize #1}}}

%
\newtheorem{Theorem}{Theorem}[section]
\newtheorem{Definition}{Definition}
\newtheorem{Proposition}{Proposition}
\newtheorem{Lemma}{Lemma}
\newtheorem{Corollary}{Corollary}
\newtheorem{Conjecture}[Theorem]{Conjecture}
\newtheorem{Property}{Property}

%
\newcommand{\ThmRef}[1]{\ref{thm:#1}}
\newcommand{\ThmLabel}[1]{\label{thm:#1}}
\newcommand{\DefRef}[1]{\ref{def:#1}}
\newcommand{\DefLabel}[1]{\label{def:#1}}
\newcommand{\PropRef}[1]{\ref{prop:#1}}
\newcommand{\PropLabel}[1]{\label{prop:#1}}
\newcommand{\LemRef}[1]{\ref{lem:#1}}
\newcommand{\LemLabel}[1]{\label{lem:#1}}
%

\newcommand \bbs{{\boldsymbol b}}
\newcommand \cbs{{\boldsymbol c}}
\newcommand \dbs{{\boldsymbol d}}
\newcommand \ebs{{\boldsymbol e}}
\newcommand \fbs{{\boldsymbol f}}
\newcommand \gbs{{\boldsymbol g}}
\newcommand \hbs{{\boldsymbol h}}
\newcommand \ibs{{\boldsymbol i}}
\newcommand \jbs{{\boldsymbol j}}
\newcommand \kbs{{\boldsymbol k}}
\newcommand \lbs{{\boldsymbol l}}
\newcommand \mbs{{\boldsymbol m}}
\newcommand \nbs{{\boldsymbol n}}
\newcommand \obs{{\boldsymbol o}}
\newcommand \pbs{{\boldsymbol p}}
\newcommand \qbs{{\boldsymbol q}}
\newcommand \rbs{{\boldsymbol r}}
\newcommand \sbs{{\boldsymbol s}}
\newcommand \tbs{{\boldsymbol t}}
\newcommand \ubs{{\boldsymbol u}}
\newcommand \vbs{{\boldsymbol v}}
\newcommand \wbs{{\boldsymbol w}}
\newcommand \xbs{{\boldsymbol x}}
\newcommand \ybs{{\boldsymbol y}}
\newcommand \zbs{{\boldsymbol z}}

\newcommand \Bbs{{\boldsymbol B}}
\newcommand \Cbs{{\boldsymbol C}}
\newcommand \Dbs{{\boldsymbol D}}
\newcommand \Ebs{{\boldsymbol E}}
\newcommand \Fbs{{\boldsymbol F}}
\newcommand \Gbs{{\boldsymbol G}}
\newcommand \Hbs{{\boldsymbol H}}
\newcommand \Ibs{{\boldsymbol I}}
\newcommand \Jbs{{\boldsymbol J}}
\newcommand \Kbs{{\boldsymbol K}}
\newcommand \Lbs{{\boldsymbol L}}
\newcommand \Mbs{{\boldsymbol M}}
\newcommand \Nbs{{\boldsymbol N}}
\newcommand \Obs{{\boldsymbol O}}
\newcommand \Pbs{{\boldsymbol P}}
\newcommand \Qbs{{\boldsymbol Q}}
\newcommand \Rbs{{\boldsymbol R}}
\newcommand \Sbs{{\boldsymbol S}}
\newcommand \Tbs{{\boldsymbol T}}
\newcommand \Ubs{{\boldsymbol U}}
\newcommand \Vbs{{\boldsymbol V}}
\newcommand \Wbs{{\boldsymbol W}}
\newcommand \Xbs{{\boldsymbol X}}
\newcommand \Ybs{{\boldsymbol Y}}
\newcommand \Zbs{{\boldsymbol Z}}

\newcommand \Absolute[1]{\left\lvert #1 \right\rvert}